\documentclass[usenatbib, useAMS]{mnras}

\usepackage{graphicx}
\usepackage{amsmath,paralist,xcolor,xspace,amssymb}
\usepackage{times}

 \newcommand{\msun}{\rm{M}_\odot}
\newcommand{\rhocr}{\rho_{\rm{cr}}}  \newcommand{\lcdm}{$\Lambda$CDM\xspace} \newcommand{\nbody}{$N$-body\xspace}
\newcommand{\eagle}{\textsc{eagle}\xspace} 
\newcommand{\gadget}{\textsc{gadget}\xspace}
\newcommand{\subfind}{\textsc{subfind}\xspace}

\title[The inner density profile of galaxy clusters]{The effect of baryons on the
  inner density profiles of rich clusters}

\author[M. Schaller et al.]  {Matthieu Schaller$^1$\thanks{E-mail:
    matthieu.schaller@durham.ac.uk}, Carlos S. Frenk$^1$, Richard G. Bower$^1$, Tom
  Theuns$^{1}$, \newauthor James Trayford$^1$, Robert A. Crain$^{2,3}$, Michelle
  Furlong$^1$, Joop Schaye$^2$,\newauthor Claudio Dalla Vecchia$^{4,5}$ and
  I. G. McCarthy$^3$\\ $^1$Institute for Computational Cosmology, Durham University,
  South Road, Durham, UK, DH1 3LE\\ $^2$Leiden Observatory, Leiden University,
  P.O. Box 9513, 2300 RA Leiden, The Netherlands\\ $^3$Astrophysics Research
  Institute, Liverpool John Moores University, 146 Brownlow Hill, Liverpool L3 5RF,
  UK\\ $^4$Instituto de Astrof\'isica de Canarias, C/ V\'ia L\'actea s/n, 38205 La
  Laguna, Tenerife, Spain\\ $^5$Departamento de Astrof\'isica, Universidad de La
  Laguna, Av. del Astrof\'isico Franciso S\'anchez s/n, 38206 La Laguna, Tenerife,
  Spain }

\begin{document}

\date{\today}

\pagerange{\pageref{firstpage}--\pageref{lastpage}} \pubyear{2014}

\maketitle

\label{firstpage}

\begin{abstract}
  We use the ``Evolution and assembly of galaxies and their environments'' (\eagle)
  cosmological simulation to investigate the effect of baryons on the density
  profiles of rich galaxy clusters.  We focus on \eagle clusters with $M_{200} >
  10^{14}\msun$ of which we have six examples. The central brightest cluster galaxies
  (BCGs) in the simulation have steep stellar density profiles, $\rho_*(r) \propto
  r^{-3}$. Stars dominate the mass density for $r < 10~{\rm kpc}$, and, as a result,
  the {\em total} mass density profiles are steeper than the Navarro-Frenk-White
  (NFW) profile, in remarkable agreement with observations. The dark matter halo
  itself closely follows the NFW form at all resolved radii ($r\gtrsim3.0~{\rm
    kpc}$). The \eagle BCGs have similar surface brightness and line-of-sight
  velocity dispersion profiles as the BCGs in the sample of \citeauthor{Newman2013a},
  which have the most detailed measurements currently available.  After subtracting
  the contribution of the stars to the central density, Newman et al. infer
  significantly shallower slopes than the NFW value, in contradiction with the \eagle
  results. We discuss possible reasons for this discrepancy, and conclude that an
  inconsistency between the kinematical model adopted by Newman et al. for their
  BCGs, which assumes isotropic stellar orbits, and the kinematical structure of the
  \eagle BCGs, in which the orbital stellar anisotropy varies with radius and tends
  to be radially biased, could explain at least part of the discrepancy.
\end{abstract}

\begin{keywords}
cosmology: theory, dark matter - galaxies: clusters: general - galaxies: haloes
\end{keywords}

\section{Introduction}
\label{sec:introduction}

Simulations of structure formation in the cold dark matter (CDM) model predict that
relaxed dark matter (DM) halos of all masses should have nearly self-similar
spherically-averaged density profiles that are well described by a simple law with a
central cusp, $\rho(r)\propto r^{-1}$, and a steeper slope, $\rho(r)\propto r^{-3}$,
at large radii \citep{Navarro1996a,Navarro1997}. This Navarro-Frenk-White (NFW)
profile provides a good approximation to halos in \nbody simulations, in which the
dark matter is treated as a collisionless fluid. Very high-resolution simulations of
this kind have shown that the profiles are not always completely self-similar and
that the inner slope could be shallower than the asymptotic NFW value
\citep{Navarro2004, Navarro2010, Neto2007, Gao2008, Gao2012,Dutton2014}. Despite
these small variations, the form of the dark matter density profile is a robust and
testable prediction of the CDM paradigm.

In the real world sufficiently massive halos contain baryons whose evolution might
affect the density structure of the dark matter.  Several processes have been
proposed that could modify the central density profile, flattening it
\citep{Navarro1996b, Pontzen2012,Martizzi2012} steepening it \citep{Blumenthal1986,
  Gnedin2004} or leaving it broadly unchanged \citep{Laporte2014}.  Understanding the
impact of these competing effects requires cosmological hydrodynamical simulations
\citep[e.g.][]{Duffy2010, Gnedin2011, DiCintio2014, Vogelsberger2014}, but these are
far more challenging than \nbody simulations and it is still unclear whether they can
treat all the relevant scales and processes sufficiently accurately.

If the effects of baryons can be reliably established, the density profiles of halos
could, in principle, reveal much about the nature of the dark matter. For example, if
the dark were self-interacting rather than effectively collisionless, with a
sufficiently large self-interaction cross-section, the inner halo density profile
could be shallower than the NFW form even in the absence of baryonic effects
\citep[e.g.][]{Spergel2000, Vogelsberger2012, Rocha2013}. Similarly, if dark matter
particles decay or annihilate, they could produce potentially detectable particles or
radiation whose intensity depends sensitively on the inner density profile.

From the observational point of view, studies of the inner dark matter density
profiles have focused on the two extremes of the halo mass distribution: dwarf
galaxies and galaxy clusters. Dwarf galaxies \citep[e.g.][]{Walker2011} are
attractive because their very high mass-to-light ratios suggest that baryonic effects
may have been unimportant. However, degeneracies in the analysis of photometric and
kinematic data have so far led to inconclusive results
\citep[e.g.][]{Strigari2010,Strigari2014}.  Galaxy clusters are also attractive
because baryons are relatively less important in the central regions than in $L_*$
galaxies and their inner profiles can be probed by strong and weak lensing, as well
as by the stellar kinematics of the central cluster galaxy.

Studies of the inner dark matter density structure in clusters have so far produced
conflicting results. For example, \cite{Okabe2013} find that a sample of $50$
clusters with good gravitational lensing data have density profiles that agree well
with the NFW form from the inner $100h^{-1}$ kpc to the virial radius. Using X-ray
observations, \cite{Pointecouteau2005}, \cite{Vikhlinin2006} and \cite{Umetsu2014}
similarly find that the total matter profile follows closely an NFW profile at
$r\gtrsim0.05R_{200} \approx 10 - 20~\rm{kpc}$.  On the other hand, combining strong
and weak lensing with stellar kinematics, \cite{Sand2004} and \cite{Newman2013a,
  Newman2013b} find that the total central profile closely follows the NFW form but,
once the contribution of the stellar component has been subtracted, the inferred dark
matter density profile is significantly flatter than NFW.

Here we analyse a sample of massive clusters ($M_{200}\gtrsim10^{14}\msun$) from the
``Evolution and assembly of galaxies and their environment'' (\eagle) cosmological
hydrodynamical simulation \citep{Schaye2014,Crain2014}. This is one of a new
generation of simulations which follow the evolution of relatively large volumes
using the best current understanding of the physical processes responsible for galaxy
formation. Since many of these processes cannot be resolved in these simulations,
they are represented by `subgrid' models which can be quite different in different
simulations \citep[e.g.][]{Schaye2010,Scannapieco2012,Okamoto2014,Vogelsberger2014}.

The \eagle simulation is sufficiently realistic that it may be compared to a range of
observed galaxy properties at different cosmic epochs.  The galaxy population in the
simulation shows broad agreement with basic properties such as the stellar mass
function and star formation history, colour, size and morphology distributions, as
well as scaling relations between photometric and structural properties
\citep{Schaye2014, Crain2014, Furlong2014,Trayford2015}.  In this
paper we focus on the effects of baryonic processes on the central density structure
of the most massive galaxy clusters in the \eagle simulation.

Our paper is organized as follows: in Section~\ref{sec:simulation}, we briefly
describe the \eagle simulation; in Section~\ref{sec:densityProfile}, we measure the
density profile of our simulated clusters; in Section~\ref{sec:slope} we focus on the
inner profile slope and compare to recent observations; in
Section~\ref{sec:measurement} we carry out a more detailed comparison with the data
of \citet{Newman2013b}.  We summarize our results in Section~\ref{sec:summary}.
Throughout this paper, we assume values of the cosmological parameters inferred from
the Planck satellite data for a \lcdm cosmology \citep{Planck2013}, the most relevant
of which are: Hubble constant, $H_0 = 67.7$~km s$^{-1}$Mpc$^{-1}$; baryon and total
matter densities in units of the critical density, $\Omega_{\rm b} = 0.0482$ and
$\Omega_{\rm m} = 0.307$ respectively, and linear power spectrum normalization,
$\sigma_8=0.829$.\\

\section{The EAGLE simulations}
\label{sec:simulation}

The \eagle set consists of a series of cosmological simulations with state-of-the-art
treatments of smoothed particle hydrodynamics and subgrid models. The simulations
reproduce the stellar mass function and other observed properties of the galaxy
population at $z=0$, and produce a reasonable evolution of the main observed galaxy
properties over cosmic time \citep{Schaye2014, Crain2014, Furlong2014}.

In brief, the largest \eagle simulation follows $1504^3\approx3.4\times10^9$ dark
matter particles and the same number of gas particles in a $100^3~\rm{Mpc}^3$ cubic
volume\footnote{Note that the units do not have factors of $h$.} from $\Lambda$CDM
initial conditions generated using $2^{\rm nd}$ order Lagrangian perturbation theory
\citep{Jenkins2010} with the linear phases taken from the public multiscale Gaussian
white noise field, \textsc{Panphasia} \citep{Jenkins2013}. The mass of a dark matter
particle is $9.7\times10^6\msun$ and the initial mass of a gas particle is
$1.8\times10^6\msun$. The gravitational softening length is $700~\rm{pc}$ (Plummer
equivalent).  The simulation was performed with a heavily modified version of the
\gadget-3 code last described by \cite{Springel2005}, using a pressure-entropy
formulation of SPH \citep{Hopkins2012} and new prescriptions for viscosity and
thermal diffusion (Dalla Vecchia (in prep.), Schaller et al. (in prep.)) and time
stepping \citep{Durier2012}.  We now summarize the subgrid model.

\subsection{Baryon physics}

The subgrid model is an improved version of that used in the \textsc{Gimic} and
\textsc{Owls} simulations \citep{Crain2009, Schaye2010}. Star formation is
implemented using a pressure-dependant prescription that reproduces the observed
Kennicutt-Schmidt star formation law \citep{Schaye2008} and uses a threshold that
captures the metallicity dependence of the transition from the warm, atomic to the
cold, molecular gas phase \citep{Schaye2004}.  Star particles are treated as single
stellar populations with a \cite{Chabrier2003} IMF evolving along the tracks provided
by \cite{Portinari1998}. Metals from AGB stars and supernovae (SNe) are injected into
the interstellar medium (ISM) following the prescriptions of \cite{Wiersma2009b} and
stellar feedback is implemented by injecting thermal energy into the gas as described
in \cite{DallaVecchia2012}. The amount of energy injected into the ISM by SNe is
assumed to depend on the local gas metallicity and density in an attempt to take into
account the unresolved structure of the ISM \citep{Schaye2014}.  Supermassive black
hole seeds are injected in halos above $10^{10}h^{-1}\msun$ and grow through mergers
and accretion of low angular momentum gas \citep{RosasGuevara2013,Schaye2014}. AGN
feedback is modelled by the injection of thermal energy into the gas surrounding the
black hole \citep{Booth2009,DallaVecchia2012}.

The subgrid model was calibrated (mostly by adjusting the intensity of stellar
feedback and the accretion rate onto black holes) so as to reproduce the present day
stellar mass function and galaxy sizes \citep{Crain2014}. The cooling of gas and the
interaction with the background radiation is implemented following
\cite{Wiersma2009a} who tabulate cooling and photoheating rates element-by-element in
the presence of UV and X-ray backgrounds \citep{Haardt2001}.

Halos were identified using the Friends-of-Friends (FoF) algorithm \citep{Davis1985}
and bound structures within them were then identified using the \subfind code
\citep{Springel2001, Dolag2009}. A sphere centred at the minimum of the gravitational
potential of each subhalo is grown until the mass contained within a given radius,
$R_{200}$, reaches $M_{200} = 200\left(4\pi\rhocr(z)R_{200}^3/3\right)$, where
$\rhocr(z)=3H(z)^2/8\pi G$ is the critical density at the redshift of interest.

\subsection{Photometry}
\label{ssec:photometry}

The luminosity and surface brightness of galaxies in the simulation are computed on a
particle-by-particle basis as described by \cite{Trayford2015}. The basic
prescription for deriving the photometric attributes of each star particle is as
follows. Each star particle is treated as a single stellar population (SSP) of the
appropriate age and metallicity as given by the simulation. The
\citet[][BC03]{Bruzual2003} population synthesis model (assuming a
\citet{Chabrier2003} IMF for consistency with the simulation) gives the integrated
spectrum of a SSP on a grid of age and metallicity.  Using bilinear interpolation we
estimate the radiated power in a particular band by integrating the spectrum through
a filter transmission curve. (Before assigning broad-band luminosities, the
metallicities are renormalised so that solar metallicity ($Z_\odot = 0.012$) is
consistent with the older solar value assumed by BC03 ($Z_\odot = 0.02$)).

Because of the limited resolution of the simulation, a star particle represents a
relatively large stellar mass. To mitigate discreteness effects, in each star
formation event star particles with stellar ages $< 100~\rm{Myr}$ are resampled from
their progenitor gas particles and the currently star-forming gas in the subhalo in
which the particle resides.  Such resampling improves the match to the observed
bimodality in galaxy colour-magnitude diagrams \citep{Trayford2015}. However,
this treatment has very little impact on the properties of the brightest cluster
galaxies (BCG) of interest here as their current star formation rates are negligible.

A modified \cite{Charlot2000} dust model is used to attenuate the light emitted by
star particles. The extinction is computed using a constant ISM optical depth and a
transient molecular cloud component that disperses after $10~\rm{Myr}$. We modified
the model so that these values scale proportionally with galaxy metallicity according
to the observed mass-metallicity relation of \cite{Tremonti2004}. The resulting
galaxy population gives a very good match to the observed luminosity function in
various commonly used broad bands \citep{Trayford2015}.

\section{The mass density profile of clusters}
\label{sec:densityProfile}

Our cluster sample consists of the six \eagle halos of mass $M_{200}> 10^{14}\msun$
(see Table~\ref{tab:haloProperties}), which we label Clusters 1 to 6. These clusters
have moderate sphericity and would likely be considered relaxed in observational
studies even if some of them fail the strict relaxation criteria used in simulations
\citep{Neto2007}.  The stellar mass function of our cluster galaxies (including the
BCG) provides a good match to observations.  Similarly, the sizes of cluster galaxies
are in good agreement with observations.  Thus, in many respects, the \eagle rich
cluster sample is quite realistic.  It is worth mentioning that \cite{Schaye2014}
showed that the gas fractions within $R_{500}$ of the clusters in our sample may be
too high when compared to observations.  However, this small disagreement does not
affect the results of this study where we focus on the very centres of the halos
($r\lesssim20~\rm{kpc}$) where the mass of gas is very small (see
Fig. \ref{fig:densityProfile}).

The main properties of our rich cluster sample are listed in
Table~\ref{tab:haloProperties}. As shown by \citet{Schaye2014}, the galaxy stellar
masses are in good agreement with abundance matching relations \citep{Moster2013}.
At the same time, the overall gas fractions within $R_{200}$ are close to the cosmic
mean, $f_{\rm b}^{\rm univ} = \Omega_{\rm b} / \Omega_{\rm m}$, as observed
\citep{Vikhlinin2006}: the AGN feedback model has succeeded in suppressing star
formation in the BCG without removing excessive amounts of gas from the halos.

\begin{table}
  \caption{Properties of the six simulated clusters studied in this work. The stellar
    mass is measured within a $30~\rm{kpc}$ spherical aperture. The baryon and
    stellar fractions are measured within $R_{200}$ and are given in units of the
    universal baryon fraction, $f_{\rm b}^{\rm{univ}} = \Omega_{\rm b} / \Omega_{\rm
      m} = 0.157$.}
\label{tab:haloProperties}
\begin{center}
\begin{tabular}{|c|c|c|c|c|c|}
Halo & $M_{200}$ & $R_{200}$ & $M_*$ & $f_{\rm b}/f_{\rm b}^{\rm{univ}}$ &
$f_*/f_{\rm b}^{\rm{univ}}$\\ & $[\msun]$ & $[\rm{kpc}]$ & $[\msun]$ & & \\ \hline
$1$ & $1.9\times 10^{14}$ & $1206$ & $4.2\times 10^{11}$ & $0.99$ & $0.07$\\ $2$ &
$3.7\times 10^{14}$ & $1518$ & $3.5\times 10^{11}$ & $0.94$ & $0.08$\\ $3$ &
$3.0\times 10^{14}$ & $1411$ & $2.9\times 10^{11}$ & $0.95$ & $0.08$\\ $4$ &
$3.1\times 10^{14}$ & $1422$ & $4.5\times 10^{11}$ & $0.97$ & $0.07$\\ $5$ &
$2.0\times 10^{14}$ & $1225$ & $2.0\times 10^{11}$ & $0.92$ & $0.08$\\ $6$ &
$2.0\times 10^{14}$ & $1229$ & $3.7\times 10^{11}$ & $0.93$ & $0.08$\\
\end{tabular}
\end{center}
\end{table}

\begin{figure}
\centering \includegraphics[width=0.9\columnwidth]{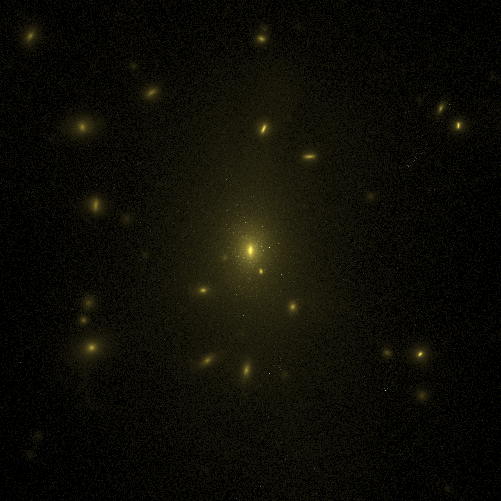}
\caption{Surface brightness map of Cluster 1, using the SDSS ugr filter system.  The
  map is $500~\rm{kpc}$ on a side and has resolution of $1~\rm{kpc}$. The central
  galaxy is easily visible and appears slightly elongated in projection. Satellite
  galaxies are also visible and cluster around the BCG.}
\label{fig:clusterMap}
\end{figure}

A surface brightness map of Cluster~1 is shown in Fig.~\ref{fig:clusterMap}. The map
is centred on the centre of potential of the halo and shows a
$500~\rm{kpc}\times500~\rm{kpc}$ area of sky. Photometry for the model galaxies in
the $u$, $g$ and $r$-band SDSS filters \citep{Doi2010} was obtained as described in
Section~\ref{ssec:photometry}. The surface brightness is then used to construct fake
colour mock images following the method of \cite{Lupton2004}. As can be seen, the
central part of the halo seems spherical. The central galaxy is slightly prolate and
the central satellite galaxies cluster around it roughly isotropically.

\subsection{The  mass density profiles of simulated halos}
\label{ssec:densityProfile}

\begin{figure*}
\includegraphics[width=\textwidth]{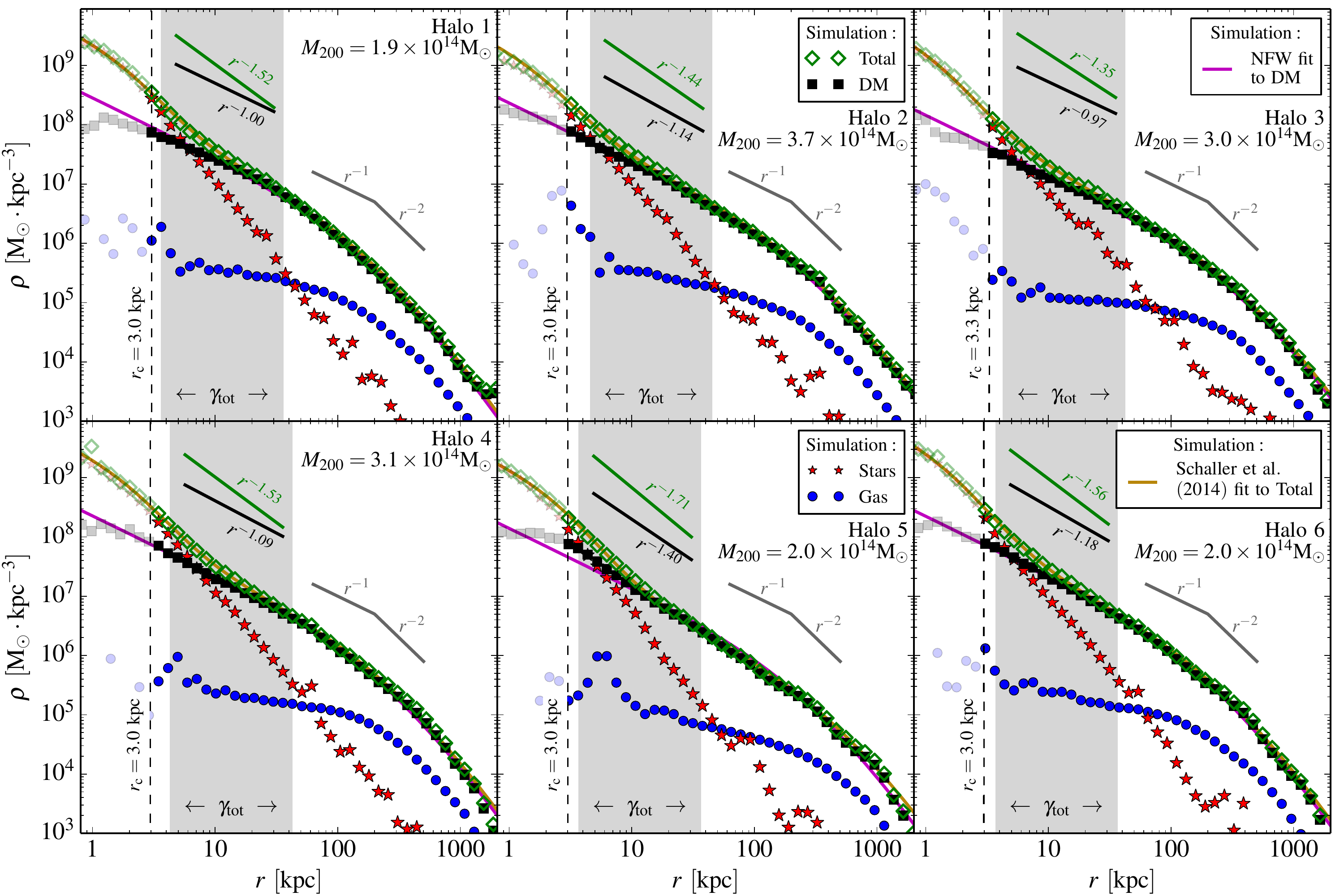}
\caption{Radial density profiles of the six simulated clusters studied in this paper
  (see Table~\ref{tab:haloProperties}). Green diamonds represent the total mass
  profile; the black squares, red stars and blue circles represent the dark matter,
  stellar and gas components respectively. The solid magenta and yellow lines are the
  best-fit NFW profile to the dark matter component and the best fit
  \citep{Schaller2014b} profile (Eqn.~\ref{eq:newProfile}) to the total mass
  distribution. The vertical dashed line in each panel shows the convergence radius,
  $r_c$, beyond which the density profile has converged to within 20\%; data points
  within this radius are shown by fainter symbols. The grey shaded regions show the
  radial range over which the logarithmic slopes, $\gamma_{\rm{tot}}$ and
  $\beta_{\rm{DM}}$, of the total and DM profiles are measured. The values of the
  slopes are given above the simulation points. The DM halos are very well fit by NFW
  profiles and the total mass profiles only deviate from NFW in the central parts
  ($r\lesssim10~\rm{kpc}$), where the stellar component dominates.  Note the
  similarities in the shapes of these six halos and the relatively small variations
  that occur mostly in the very central regions.}
\label{fig:densityProfile}
\end{figure*}

To study the density profiles of the \eagle\ clusters, we bin the particles in
logarithmically spaced radial bins centred on the minimum of the gravitational
potential. We measure the dark matter, gas and stellar components separately and then
sum all contributions to obtain the total mass profile. The result is shown in
Fig.~\ref{fig:densityProfile}, where the six panels correspond to the six clusters of
Table~\ref{tab:haloProperties}. In each panel, the green diamonds, black squares, red
stars and blue circles represent the total mass, dark matter, stellar component and
gas, respectively. The mass of each halo is indicated at the top of each panel. The
dashed vertical lines show the radius, $r_c$, above which the profile is considered
to have converged within $20\%$ \citep{Power2003, Schaller2014b}. This is a
conservative estimate of the convergence radius ($\sim3.1~\rm{kpc}$) and it is much
larger than the Plummer-equivalent softening length ($\epsilon=0.7~\rm{kpc}$) often
used as a rough estimate of the radius beyond which numerical effects become
unimportant. Data points within this `convergence radius' but at radii $r>\epsilon$
are shown using fainter symbols.

The dark matter dominates the density profiles at $r\gtrsim~8\rm{kpc}$. At smaller
radii, the stellar component dominates and exceeds the dark matter density by up to
an order of magnitude at the centre. The stellar density profiles are approximatively
constant power laws, $r^{-\alpha}$, with $\alpha\approx-3$ down to the very centre of
the galaxy. The simulation does not resolve the centre of the BCG and the slopes
measured there are probably affected by the force softening ($0.7~\rm{kpc}$ at $z=0$)
used in the \nbody solver. The peaks in the stellar components at large radii are
caused by satellites orbiting in the halo. The gas is subdominant at all radii, in
particular in the central regions where the stellar densities are almost three orders
of magnitude higher. The gas only dominates the baryon content at radii $r\gtrsim
50~\rm{kpc}$. At radii larger than $\sim 300~\rm{kpc}$, the gas profile has the same
shape as the dark matter profile. The dark matter itself has the characteristic NFW
shape, whose asymptotic behaviour is a power-law of slope $-1$ at the centre and a
power law of slope $-3$ in the outer parts.

\subsection{Fitting models to the simulated halos}
\label{ssec:fits}

The density profiles of relaxed dark matter halos in N-body simulations are well fit
by the near-universal NFW profile which has the form:

\begin{equation}
 \frac{\rho(r)}{\rhocr} = \frac{\delta_{\rm c}}{\left(r/r_{\rm
     s}\right)\left(1+r/r_{\rm s}\right)^2},
  \label{eq:nfw}
\end{equation}
\citep{Navarro1996a,Navarro1997}, where $\delta_{\rm c}$ is a characteristic
amplitude and $r_{\rm s}$, a scale length that is often expressed in terms of the
concentration, $c_{200}= R_{200}/r_{\rm s}$. Both $\delta_{\rm c}$ and $c_{200}$
correlate with halo mass, $M_{200}$, so the NFW profile is fully specified by the
halo mass.  In our simulations, the cold gas and stars, which contribute only a small
fraction of the total mass, are concentrated towards the centre, while the hot gas
beyond the central regions closely follows the dark matter profile.  Thus, even in
the presence of baryons, the \emph{dark matter} still closely follows an NFW
profile. In the case of halos of mass $M_{200} \sim 10^{12}-10^{13}\msun$, the
profile is slightly modified in the centre by a modest contraction due to the
presence of stars \citep{Duffy2010, DiCintio2014, Schaller2014b}.

Baryon contraction is less important in halos of mass $M_{200} \sim 10^{14}\msun$
which are well fit by an NFW profile, as can be seen in
Fig.~\ref{fig:densityProfile}, where the solid magenta line shows the best-fit NFW
profile. The fit was performed using all radial bins from the resolution limit,
$r_{\rm{c}}\sim3~\rm{kpc}$, to the virial radius, $R_{\rm{vir}}\sim2\rm{Mpc}$. We
have checked that the best fitting parameter values are largely insensitive to the
exact radial range used, provided that both the $\rho(r)\rightarrow r^{-1}$ and
$\rho(r) \rightarrow r^{-3}$ regimes of the profile are well sampled.  In all but one
case, the magenta line closely tracks the DM profile plotted as black squares. The
exception is halo $5$ (bottom row, middle panel) which shows a slight deviation from
the NFW form in the radial range $3-7~\rm{kpc}$, where some contraction is seen,
possibly as a result of the recent accretion of a large substructure. The best-fit
NFW parameters are listed in Table~\ref{tab:NFWfit}. The mean and scatter in
concentration (nearly a factor of 2) of our halos are consistent with the results
obtained for relaxed halos in the Millennium simulation by \citet{Neto2007}, who
found a concentration, $c_{200}=4.51_{-0.62}^{+0.71}$, for halos of mass
$M_{200}=10^{14}\msun$.

\begin{table}
\caption{Parameters of the best-fit NFW profiles (Eqn.~\ref{eq:nfw}) to the
  \emph{dark matter} component of our halos.}
\label{tab:NFWfit}
\begin{center}
\begin{tabular}{|c|c|c|c|c|c|}
Halo & $M_{200}$ & $R_{200}$ & $r_{\rm s}$ & $c_{200}$ & $\delta_{\rm c}$\\ &
$[\msun]$ & $[\rm{kpc}]$ & $[\rm{kpc}]$ & & \\ \hline $1$ & $1.9\times10^{14}$ &
$1206$ & $199.2$ & $6.1$ & $1.1\times10^{4}$\\ $2$ & $3.7\times10^{14}$ & $1518$ &
$350.8$ & $4.3$ & $5.3\times10^{3}$\\ $3$ & $3.0\times10^{14}$ & $1411$ & $452.0$ &
$3.1$ & $2.5\times10^{3}$\\ $4$ & $3.1\times10^{14}$ & $1422$ & $305.1$ & $4.7$ &
$5.9\times10^{3}$\\ $5$ & $2.0\times10^{14}$ & $1225$ & $331.1$ & $3.7$ &
$3.3\times10^{3}$\\ $6$ & $2.0\times10^{14}$ & $1229$ & $245.9$ & $5.0$ &
$7.2\times10^{3}$\\
\end{tabular}
\end{center}
\end{table}

While the dark matter is well described by an NFW profile, the \emph{total matter}
profile in our halos is not. In our study of the entire halo population in the
\eagle\ simulation, we introduced the following fitting formula for the total matter
\citep{Schaller2014b}:

\begin{equation}
 \frac{\rho(r)}{\rhocr} = \frac{\delta_{\rm c}'}{\left({r}/r_{\rm s}'\right)\left(1 +
   {r}/{r_{\rm s}'}\right)^2} + \frac{\delta_{\rm i}}{\left({r}/{r_{\rm
       i}}\right)\left(1+\left({r}/{r_{\rm i}}\right)^2\right)}.
\label{eq:newProfile}
\end{equation}
The first term has the NFW form and describes the overall shape of the profile; the
second term is a correction that reproduces the stellar cusps ($\rho_*\propto
r^{-3}$), together with any dark matter contraction due to the presence of
baryons. The dark yellow solid lines in the six panels of
Fig.~\ref{fig:densityProfile} show the best-fit profiles of this kind to each halo
which, as may be seen from the figure, represent the data well over the entire
resolved radial range. The best-fit parameter values are listed in
Table~\ref{tab:NewProfilefit}.

\begin{table}
  \caption{Parameters of the best-fit profiles of the form of
    Eqn.~\ref{eq:newProfile} \citep{Schaller2014b} to the \emph{total} matter
    distribution in our halo sample. }
\label{tab:NewProfilefit}
\begin{center}
\begin{tabular}{|c|c|c|c|c|c|}
Halo & $M_{200}$ & $r_{\rm s}'$ & $\delta_{\rm c}'$ & $r_{\rm i}$ & $\delta_{\rm i}$
\\ & $[\msun]$ & $[\rm{kpc}]$ & & $[\rm{kpc}]$ &\\ \hline $1$ & $1.9\times10^{14}$ &
$209.3$ & $1.2\times10^{4}$& $ 2.1$ & $8.8\times10^{6}$\\ $2$ & $3.7\times10^{14}$ &
$369.8$ & $5.5\times10^{3}$& $ 2.0$ & $6.4\times10^{6}$\\ $3$ & $3.0\times10^{14}$ &
$433.8$ & $3.2\times10^{3}$& $ 1.4$ & $1.1\times10^{7}$\\ $4$ & $3.1\times10^{14}$ &
$321.6$ & $6.2\times10^{3}$& $ 2.2$ & $7.2\times10^{6}$\\ $5$ & $2.0\times10^{14}$ &
$529.1$ & $1.5\times10^{3}$& $ 2.7$ & $4.0\times10^{6}$\\ $6$ & $2.0\times10^{14}$ &
$277.0$ & $6.4\times10^{3}$& $ 1.6$ & $1.5\times10^{7}$\\
\end{tabular}
\end{center}
\end{table}

\section{The inner density profile}
\label{sec:slope}

A testable prediction from simulations evolving only dark matter of the $\Lambda$CDM
model is that the average slope of the inner mass profile ($r\ll r_{\rm s}$) should
tend to the NFW value of $-1$. Steeper profiles might be explained by baryon effects
causing some contraction. Significantly shallower profiles in massive halos, on the
other hand, would be more difficult to explain. Explosive baryon effects could lower
the inner dark matter density, and even induce cores, but only in dwarf galaxies
\citep{Navarro1996b,Read2005,Pontzen2014}. In massive halos \citet{Martizzi2012} have
argued that AGN feedback could introduce small ($\sim 10~\rm{kpc}$) cores, but it is
unclear if this kind of feedback is compatible with the observed stellar masses of
BCGs and the baryon fractions of clusters.  Shallower inner profiles could also be
generated if the dark matter is self-interacting \citep{Vogelsberger2012,Rocha2013}.

\subsection{Total mass profiles: simulation results}
\label{ssec:simulation_slope}

A quantity that can be derived from observational data in selected samples of rich
clusters is the \emph{average logarithmic slope} of the inner density profile of the
total mass, that is dark matter and baryons \citep[e.g.][]{Sand2004, Newman2013a}:

\begin{equation}
 \gamma_{\rm{tot}} \equiv -\left\langle \frac{d\log\rho_{\rm{tot}}(r)}{d\log r}
 \right\rangle_{r\in[0.003R_{200}, 0.03R_{200}]},
\end{equation}
where the average is over the radial range $[0.003R_{200}, 0.03R_{200}]$.  It is
important to recognize that the radial range typically probed by the data is {\it
  not} the asymptotic regime, $r\rightarrow0$, where the NFW profile tends to
$\rho(r)\propto r^{-1}$. Instead, in the region probed by observations, the NFW
formula (Eqn.~\ref{eq:nfw}) predicts values of the inner slope significantly steeper
than $-1$ (i.e. $\gamma_{\rm{tot}} > 1$):

\begin{equation}
\gamma_{\rm{tot}} = 1 + \log_{10}\left( \frac{ \left( 1 + 0.03c_{200} \right)^2 } {
  \left(1 + 0.003c_{200} \right)^2 } \right),
\label{eq:NFW_slope}
\end{equation}
which, for the expected range of cluster concentrations ($c_{200}\in [3,5]$), gives
$\gamma_{\rm{tot}}\approx 1.1$.

\begin{figure}
\includegraphics[width=\columnwidth]{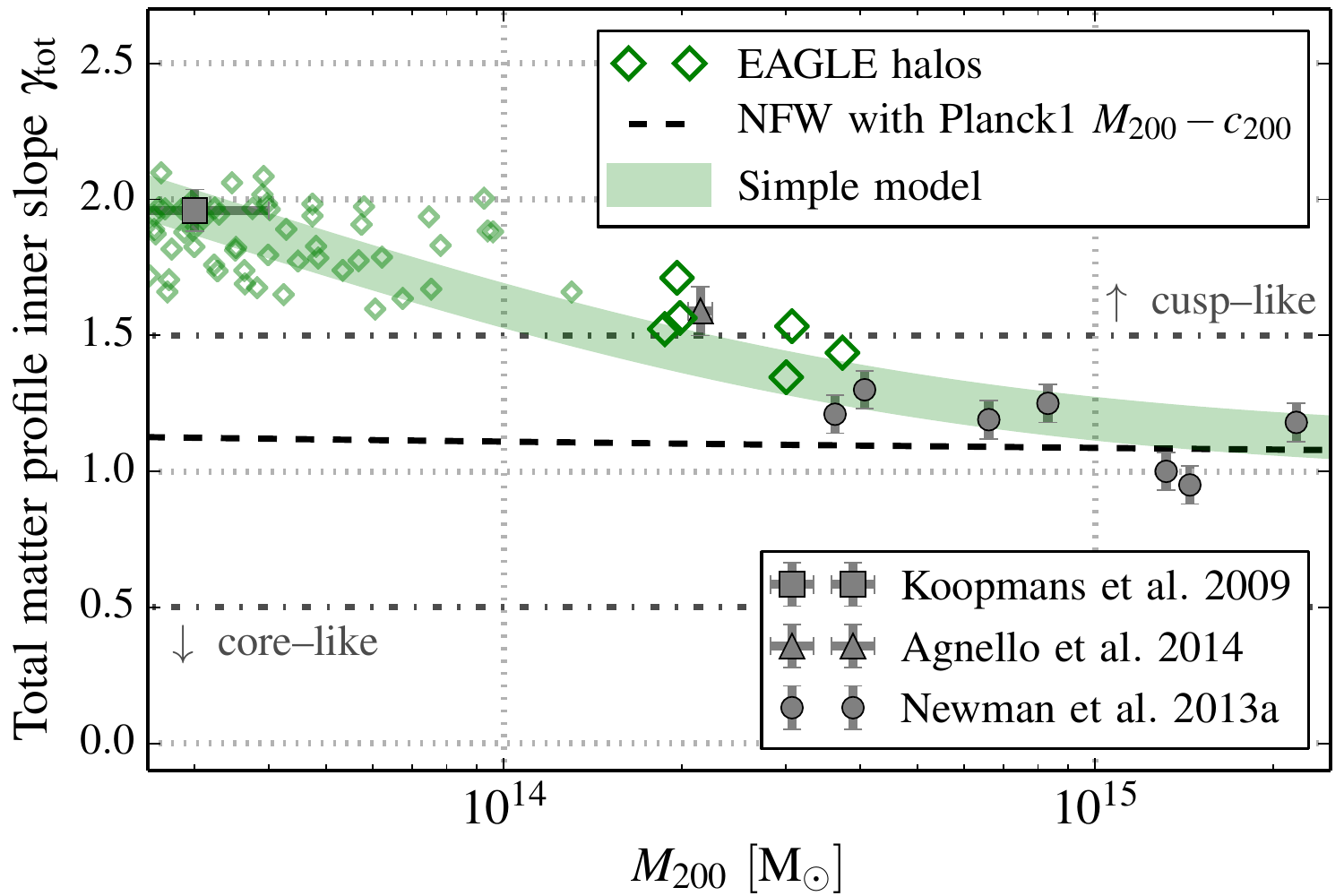}
\caption{The logarithmic slope of the inner density profile of the {\em total} mass
  distribution, $\gamma_{\rm{tot}}$, as a function of halo mass, $M_{200}$. The
  dashed line shows the average slope of the NFW profile over the range in which
  $\gamma_{\rm{tot}}$ is defined. The large green diamonds represent the six
  \eagle\ halos in our sample and the small green diamonds the smaller mass
  \eagle\ halos.  The grey symbols with error bars are the slopes measured by
  \citet{Koopmans2009} for 58 early type galaxies in the SLACS survey (square), the
  slope inferred by \citet{Agnello2014} from globular clusters orbits in M87
  (triangle) and the slopes measured by \citet{Newman2013a} for seven massive
  clusters (circles). As a guide, the grey dash-dotted lines lines demarcate slopes
  that may be construed as ``core-like'' and ``cuspy''. The lower mass
  \eagle\ clusters are cuspier than an NFW halo because of the contribution of the
  stellar component which, however, becomes increasingly less important for larger
  mass halos. The \citet{Newman2013a} data lie along the extrapolation of the trend
  seen in the \eagle\ clusters.}
\label{fig:slope_total}
\end{figure}

The radial range over which $\gamma_{\rm{tot}}$ is typically measured in
observational studies is shown for our clusters as a grey shaded region in each panel
of Fig.~\ref{fig:densityProfile}. The values of the slope predicted by our
simulations in this range are shown above the data points.  Values of
$\gamma_{\rm{tot}}$ for our halo sample are plotted as a function of halo mass in
Fig.~\ref{fig:slope_total} (large green diamonds), which also includes data for halos
in the simulation volume that are less massive than those in our main sample (small
green diamonds).  The black dashed line shows the slope of the NFW profile obtained
from Eqn.~(\ref{eq:NFW_slope}) and the mass-concentration relation of
\citet{Schaller2014b}. As a guide, we include two dashed-dotted lines showing where
profiles may be considered to be ``cuspy'' ($\gamma_{\rm{tot}} > 1.5$) or
``core-like'' ($\gamma_{\rm{tot}} < 0.5$). The exact position of these lines is, of
course, arbitrary.

The high mass tail of the cluster population is not represented in the limited volume
of the \eagle\ simulation. However, the general behaviour of massive clusters can be
readily inferred from the trends seen for smaller halos.  A halo of mass
$M_{200}\approx2\times10^{15}\msun$ has $R_{200}\approx2~\rm{Mpc}$ and thus
$\gamma_{\rm tot}$ is centred (logarithmically) around $r=20~\rm{kpc}$. In this
region the profile is dominated by dark matter, even in the case of large, extended
galaxies. Thus, $\gamma_{\rm tot}$ is unaffected by the BCG and directly reflects the
slope of the DM profile which \cite{Schaller2014b} showed has a slope close to or
slightly steeper than the NFW value, as given by eq.~\ref{eq:NFW_slope}. Thus, for
halos of $M_{200}\approx2\times10^{15}\msun$ we expect $\gamma_{\rm
  tot}\approx1.1$. This conclusion is consistent with the collisionless model of
\cite{Laporte2014} who also find slopes close to the NFW value for
$M_{200}\sim10^{15}\msun$ clusters.

Based on this argument we can construct a simple model, consistent with the results
for low mass halos, to extrapolate the slopes measured for the \eagle\ clusters into
the mass range appropriate to rich clusters. This model is designed to capture the
general behaviour of $\gamma_{\rm tot}$ on mass scales larger than
$M_{200}\approx2\times10^{13}\msun$.  It assumes that the total matter profile is
made up of an NFW dark matter halo plus a stellar component which, in order to be
consistent with relevant observational analyses, we take to be a ``dual pseudo
isothermal elliptical mass distribution'' (dPIE) \citep{Eliasdottir2007}. The value
of the halo mass determines the concentration of the halo and we infer the stellar
mass of the central galaxy from abundance matching \citep[e.g.][]{Moster2013}. For
the dPIE profile, we adopt the mean scale radius and core radius of the best-fitting
profiles for our BCGs and keep them fixed while varying the normalisation to match
the stellar mass of interest.  (We verified that varying the values of the parameters
of this model does not affect our results.) In this way we construct the total mass
profile and measure its slope, which we show as the green band in
Fig.~\ref{fig:slope_total}.  The slopes of the total mass profiles of the largest
\eagle\ clusters plotted in Fig.~\ref{fig:densityProfile} are slightly steeper than
the NFW value over the radial range over which $\gamma_{\rm{tot}}$ is defined. This
mostly reflects the contribution of stars to the inner matter density.  By contrast,
the values inferred for more massive clusters are closer to the NFW value.

We now turn to the slope of the dark matter profiles. Unlike the total mass profile,
the dark matter profile cannot be measured directly from observations, but must
instead be inferred through detailed modelling, which requires a number of
assumptions. The dark matter profile in the simulations can, of course, be directly
measured and the simulations can be used to test the consistency of the assumptions
required in the modelling of the observational data.

\begin{figure}
\includegraphics[width=\columnwidth]{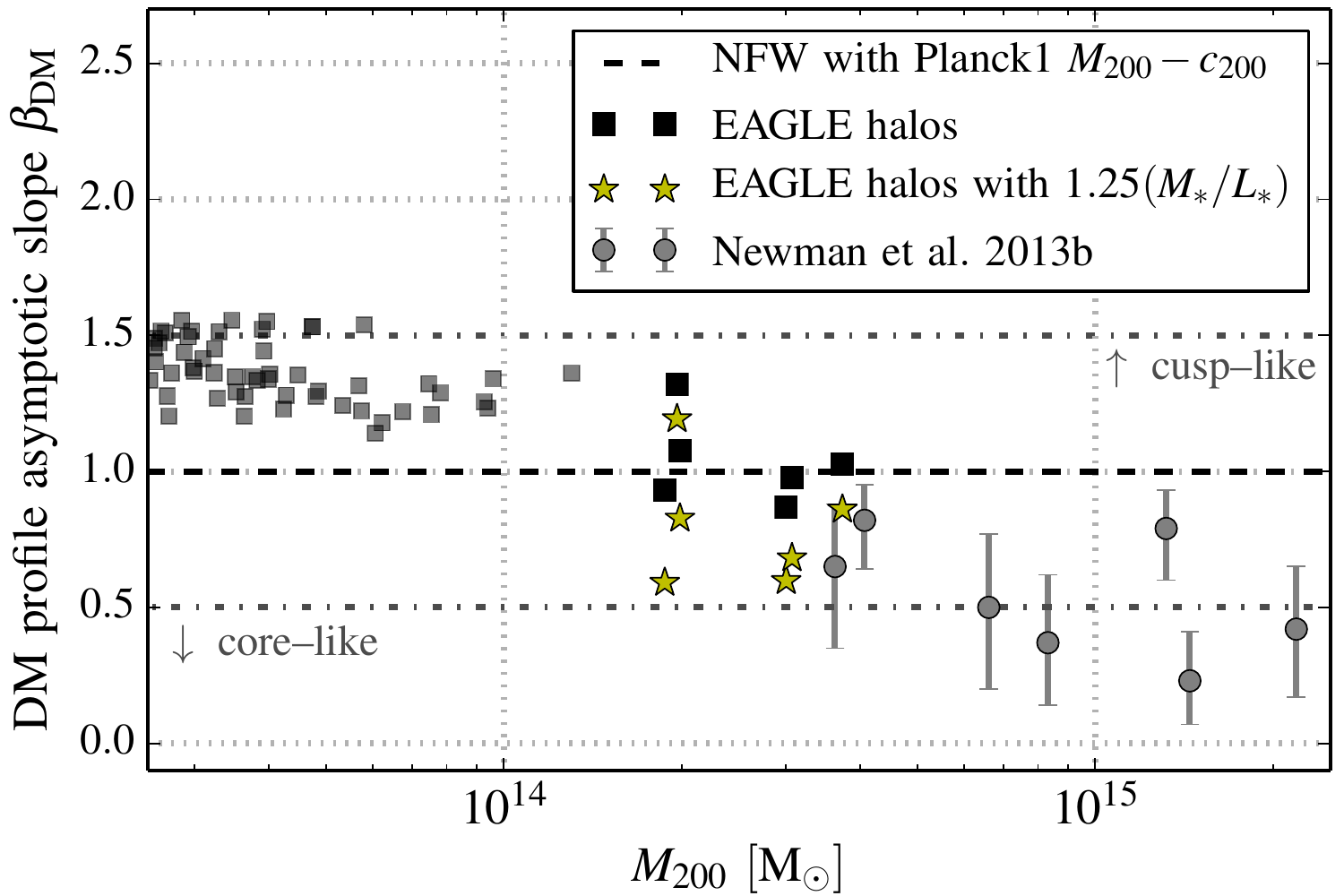}
\caption{The asymptotic logarithmic slope of the inner dark matter density profile,
  $\beta_{\rm{DM}}$, as a function of halo mass, $M_{200}$. The dashed line shows the
  NFW value of $-1$.  The large black squares show the values measured for the six
  massive clusters in our \eagle\ sample and the small black squares those measured
  for smaller \eagle\ clusters. The yellow stars show the slopes that would be
  inferred for our sample if the stellar mass-to-light ratio is overestimated by
  $25\%$ (see Section \ref{ssec:M_L} for details).  The grey circles with error bars
  are the values inferred by \citet{Newman2013b}.  As in Fig.~\ref{fig:slope_total},
  the grey dash-dotted lines lines demarcate slopes that may be construed as
  ``core-like'' and ``cuspy''. }
\label{fig:slope_DM}
\end{figure}

The average slope of the dark matter density profile over the same radial range used
to define $\gamma_{\rm{tot}}$ ($[0.003R_{200}, 0.03R_{200}]$) for our sample of
simulated clusters is indicated by the black line in the grey shaded regions in
Fig.~\ref{fig:densityProfile} . Some observational analyses attempt to constrain the
asymptotic slope, $\beta_{\rm{DM}}$, of a generalized NFW profile (gNFW):

\begin{equation}
 \frac{\rho_{\rm{gNFW}}(r)}{\rhocr} = \frac{\delta_{\rm c}}{(r/r_{\rm
     s})^{\beta_{\rm{DM}}}(1+r/r_{\rm s})^{3-\beta_{\rm{DM}}} }.
 \label{eq:gNFW}
\end{equation}
This profile is often used to quantify deviations from the NFW form to which it
reduces for $\beta_{\rm{DM}} = 1$ (Eqn. \ref{eq:nfw}). We fit this profile to the
dark matter of our simulated halos and plot the resulting values of $\beta_{\rm{DM}}$
as a function of halo mass, $M_{200}$, in Fig.~\ref{fig:slope_DM}, which is the dark
matter analogue of Fig.~\ref{fig:slope_total}. As may be seen, the \eagle\ clusters
(black squares) have inner slopes consistent with the NFW expectation
(Eqn.~\ref{eq:NFW_slope}). As was the case for the total matter profile, the inner
dark matter profile slopes also show significant scatter, with $\beta_{\rm{DM}}$
varying by as much as $\sim 0.4$ for halos of similar mass.

\subsection{Total mass profiles: overview of recent observational data}
\label{ssec:observed_slope}

By combining different observational techniques, the \emph{total} matter profile of
clusters can be estimated. Two techniques have been used to probe the central mass
distributions: strong lensing and modelling of the orbits of globular clusters (GC)
or BCG stars. The former relies on a chance alignment of the cluster with a
background galaxy and is, by nature, rare since only a few galaxy clusters present
strong lensing arcs at the radii of interest. Similarly, the use of globular cluster
(GC) orbits as tracers of the potential is limited to clusters that are close enough
for the GCs to be unambiguously detected.  Stellar velocity dispersion measurements
of the central galaxy can also be used to constrain the mass near the centre of the
halo but high-resolution spectroscopy is required. We will now compare our simulated
cluster slopes to recent observational data. Although there is a wealth of data
available for profiles at large radii, we focus exclusively on the inner regions
which are the most sensitive to the nature of the dark matter.

The grey square with error bars at the low mass end of Fig.~\ref{fig:slope_total}
shows the average slope measured for 58 early type galaxies in the SLACS survey by
\cite{Koopmans2009} using a combination of strong lensing and stellar velocity
dispersion measurements. Our simulation agrees perfectly with this data point. At the
more massive end, \cite{Newman2013a} derived total mass profiles,
$\rho_{\rm{tot}}(r)$, from projected mass profiles, which they estimated using strong
and weak lensing data, together with the surface brightness and resolved stellar
kinematics of the BCGs in a sample of seven clusters.  The grey circles with error
bars in Fig.~\ref{fig:slope_total} show their results. Five of these clusters have
higher masses than the largest clusters in the relatively small \eagle\ simulation
volume, but the two lightest ones fall in the region represented in our
simulation. Their values of $\gamma_{\rm{tot}}$ agree very well with those measured
directly in our simulated clusters while the values for the more massive five lie in
the region predicted by the simple model used to extrapolate the \eagle{} results
described in Section~\ref{ssec:simulation_slope}.
 
An independent measurement of the total inner density profile which does not rely on
lensing data was carried out by \cite{Agnello2014} using the orbits of globular
clusters in the halo of M87. The slope, $\gamma_{\rm{tot}}$ , inferred from their
best fitting broken power-law model is shown as a triangle with error bar In
Fig.~\ref{fig:slope_total}. This data point also agrees extremely well with the
results for the \eagle clusters.

We conclude that the inner density profiles of the {\em total} mass distribution in
the \eagle{} clusters are in good agreement with the best current data
\citet{Koopmans2009, Newman2013a, Agnello2014}.  In both simulated and observed
clusters, the inner profile slopes exhibit considerable scatter reflecting the
variety of factors that affect the density structure, such as halo assembly history,
shape and substructure distribution, BCG star formation and merger history, etc.

\subsection{Dark matter density profiles}

The situation is more complicated for the density profile of the dark matter since
this is not directly accessible to observations. Instead, this profile must be
inferred from a model to disentangle the contributions of the dark and visible
components from the measured total mass profile. Wide radial coverage is needed fully
to sample the two components and effect the decomposition.  Strong lensing data
seldom sample the range, $r\lesssim10~\rm{kpc}$, where the influence of baryons
starts to play a role and so lensing data need to be supplemented by, for example,
kinematical data for the stars of the BCG. Such data exist for only a handful of
clusters \citep[e.g.][]{Sand2004, Newman2013a}. The study by \cite{Newman2013a} is
particularly interesting by virtue of the quality of the data and the comprehensive
analysis performed. In the remainder of this paper we will therefore focus on the
comparison with these data.

The model assumed by \cite{Newman2013a} is a generalized NFW (gNFW) profile for the
dark matter and a dPIE profile for the galaxy.  The authors estimated the parameters
values that minimize the difference between the model and the inferred lensing mass,
the measured profiles of stellar velocity dispersion, $\sigma_{\rm{l.o.s.}}$, and
surface brightness, $S$.  In addition to the parameters describing the dark matter
profile, the minimization procedure also constrains the stellar mass-to-light ratio,
$\Upsilon_*$. This is an important parameter since, at a given radius, it is
degenerate with the dark matter mass: one can always trade dark matter for unseen
stellar mass at that radius,
\begin{equation}
 \label{eq:measured_rho}
 \rho_{\rm{DM}}(r) = \rho_{\rm{tot}}(r) - \Upsilon_* \times S(r).
\end{equation}
This degeneracy can be broken by measuring the total density and surface brightness
as a function of projected radius, $R$, and assuming that the stellar mass-to-light
ratio is constant.

The values of $\beta_{\rm{DM}}$ (Eqn. \ref{eq:gNFW}) inferred by \cite{Newman2013b}
for their sample of clusters are shown as grey circles in
Fig.~\ref{fig:slope_DM}. The error bars indicate the 16th and 84th percentiles of the
posterior distribution of $\beta_{\rm{DM}}$ returned by their model (not including
systematics). These lie well below the values for the \eagle\ clusters (black
squares) and are clearly inconsistent with them given the quoted errors.  From our
earlier discussion it seems unlikely that the discrepancy can be due to the slightly
smaller masses of the \eagle\ clusters compared to those in the observed sample,
since the \eagle\ clusters have dark matter inner slopes that are either close to or
slightly steeper (due to contraction) than the NFW value. Thus, we conclude that
profile slopes as shallow as those inferred by \cite{Newman2013b} are not present in
$\Lambda$CDM simulations with the baryon physics modelled in \eagle. This conclusion
is surprising since the {\em total} mass profiles of the real and simulated clusters
agree remarkably well.  We will now discuss possible reasons for this apparent
discrepancy.

\section{Discussion}
\label{sec:measurement}

We saw in the preceding section that the inner slopes of the density profiles of the
dark matter halos in the \eagle\ clusters differ from the profiles inferred by
\cite{Newman2013b} for their sample of seven clusters. There are several possible
explanations for the discrepancy. One is that the simulations do not model the
correct physics. This would be the case if the dark matter does not consist of cold
collisionless particles but of particles that undergo self-interactions
\citep[e.g.][]{Spergel2000, Vogelsberger2012,Rocha2013}. Cluster simulations would be
required to determine whether the slopes found by Newman et al. can be explained for
reasonable values of the self-interaction cross-section and a reasonable model for
the baryonic physics.

Another possibility is that the dark matter is indeed cold and collisionless but the
injection of energy from an AGN has flattened the NFW cusp. This is a scaled-up
version of the mechanism originally invoked by \cite{Navarro1996b} to explain the
possible existence of cores in dwarf galaxies. The simulations of \cite{Martizzi2012}
show precisely this effect, but kiloparsec-scale cores are only produced by injecting
very large amounts of AGN energy into the surrounding gas. Our simulations have
weaker AGN feedback, as required to achieve a good match to the massive end of the
observed stellar mass function, and do not produce cores. It is unclear if feedback
as intense as that required by \cite{Martizzi2012} would lead to a similarly good
match to the global properties of the galaxy population \citep{Schaye2014,Crain2014,
  Furlong2014}.

The disagreement between the inner dark matter profiles of the \eagle\ clusters and
of the clusters in the \cite{Newman2013b} sample could also be due to a mismatch
between the directly observable quantities, $\sigma_{\rm{l.o.s.}}(R)$ and $S(R)$, and
the corresponding quantities for the \eagle clusters, or to systematic effects either
in the selection of the observational sample or in the method used to inferred the
inner dark matter slopes. We will now discuss these possibilities.

The most direct way to carry out the comparison would be to replicate the analysis of
\cite{Newman2013a} on our simulated clusters. Unfortunately, the exact model and
fitting pipeline used by them is not available to us and, as we will see below, the
results are very sensitive to small changes in the assumption of the analysis
pipeline. We therefore restrict our comparison to directly observable quantities and
discuss how some of the assumptions made could impact the inferred values of
$\beta_{\rm{DM}}$.

\subsection{Surface brightness profiles}
\label{ssec:surfaceBrightness}

Stars are the dominant contributors to the density in the central regions of the
\eagle\ clusters and probably also in the real data. Clearly, if the surface
brightness of the simulated clusters differed significantly from the observations,
subtraction of this component could lead to different results for the slope of the
dark matter profile in the two cases.

As discussed in Section~\ref{ssec:photometry}, the luminosity of each stellar
particle in the simulations is obtained from a \cite{Bruzual2003} population
synthesis model assuming the \cite{Chabrier2003} IMF. To compare our halos with
observations, we derive magnitudes in the four HST filters (F606W, F625W, F702W and
F850LP) used by \cite{Newman2013a}. We placed our clusters at $z=0.25$, the mean
redshift of that sample, by redshifting the spectra before applying the HST
filters\footnote{Effectively applying a reverse K-correction.}  and dimming the
luminosities by a factor $(1+z)^{-4}$. To account for the somewhat smaller masses of
the \eagle\ clusters compared to those in the sample of Newman et al. (whose mean
mass is $M_{200}=1.03\times 10^{15}\msun$) we scaled up their surface brightnesses by
a modest factor, $(M_{200}/1.03\times 10^{15}\msun)^{1/6}$, derived assuming that the
luminosity $L\propto M_{200}^{1/2}$ and that the stellar density remains constant in
the narrow range or relevant halo masses\footnote{Note that this is a more
  conservative rescaling factor than simply assuming $L\propto M_{200}$.}  We then
chose 10,000 random lines-of-sight through each cluster and projected the particles
along those axes onto the plane of the (virtual) sky. Finally, we binned the
particles radially from the centre of the potential to derive the stellar surface
brightness.

\begin{figure*}
\includegraphics[width=\textwidth]{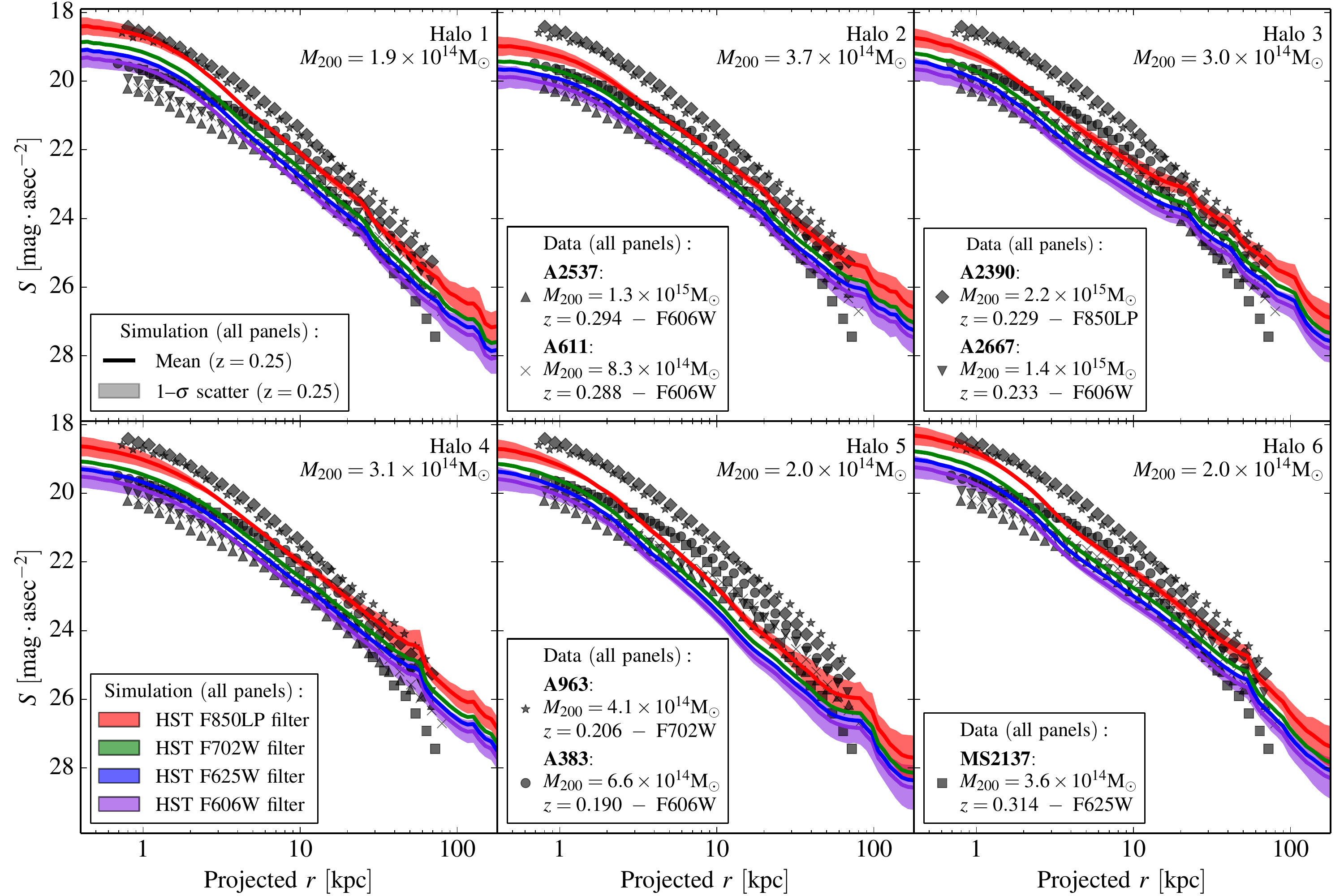}
\caption{Surface brightness profiles of the six \eagle\ clusters in our sample
  (placed at $z=0.25$) in the four HST filters (from top to bottom: F850LP, F702W,
  F625W and F606W) used by \citet{Newman2013a} in AB magnitudes per
  $\rm{arcsec}^{-2}$. The solid lines show the mean profile scaled by the factor,
  $(M_{200}/1.03\times 10^{15}\msun)^{1/6}$, averaged over 10,000 random
  lines-of-sight. The shaded regions show the $1\sigma$ scatter for the reddest and
  bluest filters. (The other filters have similar scatter.) The black symbols show
  the measured surface brightness profiles of the seven clusters observed by
  \citet{Newman2013a} whose redshifts are given in the legend together with the
  filter used. The clusters in the simulations have surface brightness profiles in
  reasonable agreement with those observed. }
\label{fig:surfaceBrightness}
\end{figure*}

The surface brightness profiles of our six \eagle\ clusters are plotted in
Fig.~\ref{fig:surfaceBrightness}. The solid lines show the mean profiles averaged
over 10,000 lines-of-sight in the four different HST filters and the shaded regions
the $1\sigma$ scatter around these values. The black symbols correspond to the
measurements taken from \cite{Newman2013a} with physical radii derived from their
angular sizes and redshift measurements.  Although the \eagle\ clusters have a
slightly smaller total mass, the surface brightness of their central galaxies are in
quite good agreement with those of the \citeauthor{Newman2013a} sample: the shapes of
the profiles are somewhat different, with our clusters having a slightly shallower
inner slope than the observed clusters.

A striking feature of Fig.~\ref{fig:surfaceBrightness} is the small scatter in the
simulations for the different lines-of-sight. Near the centre the scatter is
dominated by the presence of foreground satellites rather than by the orientation of
the BCG.  Another interesting feature is the large object-to-object variation, both
in the simulations, where the central luminosities vary by around $0.8~\rm{mag}$, and
in the observations, where the variation is even larger, almost $2~\rm{mag}$, with no
apparent correlation with halo mass. Our simulated halos lie well within the
observational scatter but themselves show somewhat smaller scatter.  At large radii,
5 out of our 6 clusters appear to be slightly more luminous than the real clusters.
However, this is the region where the observational data terminate and where
background subtraction becomes significant.

We conclude that the surface brightness profiles of the \eagle\ clusters are
sufficiently similar to those of the \cite{Newman2013b} sample that differences in
the starlight distribution cannot be the reason for the discrepancy between the dark
matter profiles in the simulations and those inferred from the data.

\subsection{Velocity dispersion profiles}
\label{ssec:velocityDispersion}

\begin{figure*}
\includegraphics[width=\textwidth]{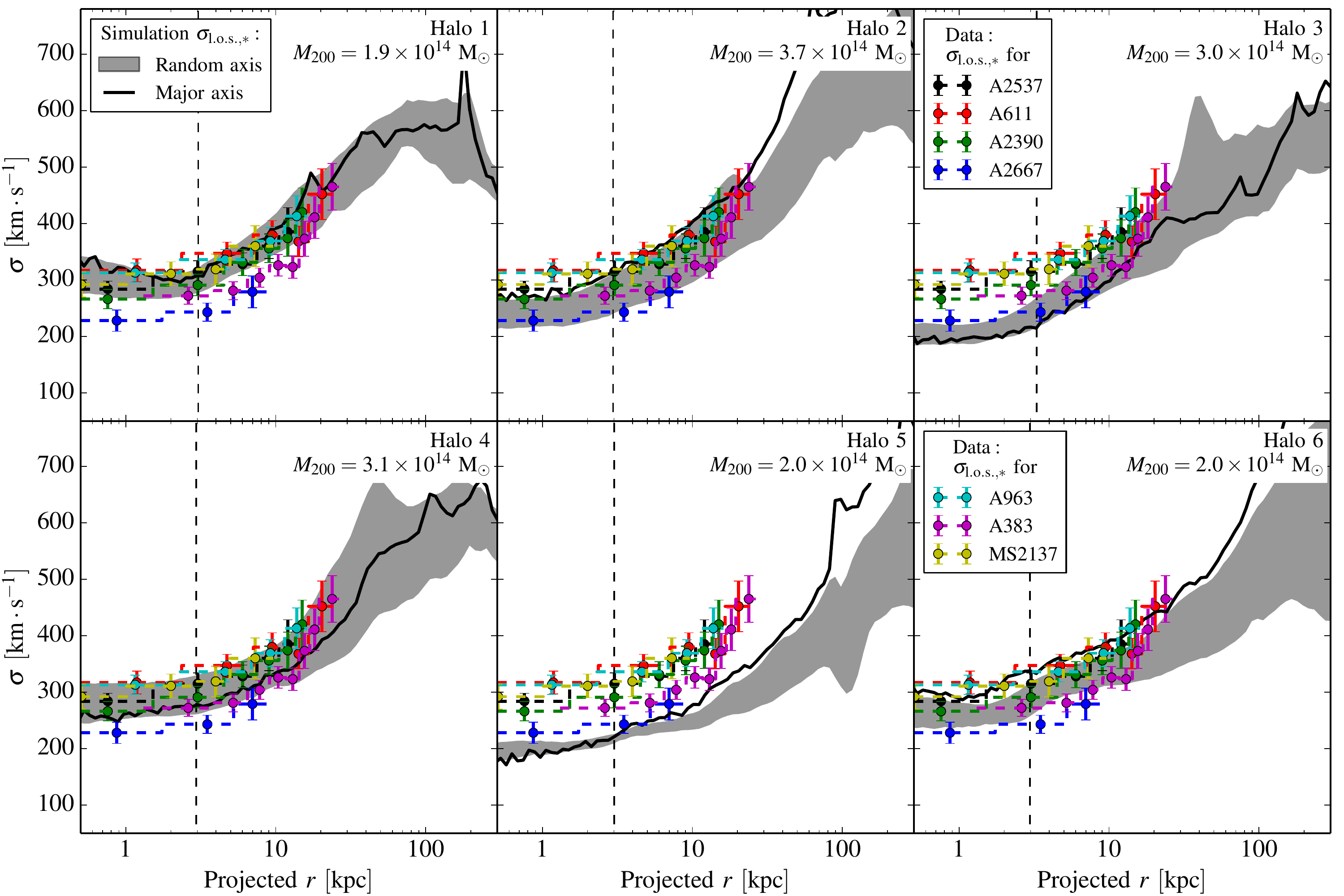}
\caption{Stellar velocity dispersion along the line-of-sight as a function of
  projected radius for the six \eagle clusters listed in Table~
  \ref{tab:haloProperties}. The dispersions have been rescaled by a factor
  $(M_{200}/1.03\times 10^{15}\msun)^{1/6}$ to correct for the slightly lower masses
  of these clusters compared to the mean of the observational sample of
  \citet{Newman2013a}. The grey shaded region is the $1\sigma$ scatter obtained when
  looking at the halos from 10,000 random lines-of-sight. The black solid line is the
  profile as seen from a line-of-sight oriented along the galaxy's major axis.  The
  vertical dashed line on each panel shows the 3D convergence radius, $r_c$. The
  coloured dashed lines with error bars are the measurements for the seven clusters
  observed by \citet{Newman2013a}.  In three of the six \eagle halos the velocity
  dispersion profile measured along the major axis is biased high.}
\label{fig:velocityDispersion}
\end{figure*}

The line-of-sight stellar velocity dispersion, $\sigma_{\rm l.o.s.}$, of our six
halos as a function of projected radius is shown in
Fig.~\ref{fig:velocityDispersion}.  Since the \eagle clusters are less massive than
the clusters in the sample of \cite{Newman2013a}, in order to facilitate a
comparison, the velocity dispersions of the simulated clusters have been rescaled, as
before, to the mean mass of the Newman et al. sample by multiplying the velocity
dispersions by the corresponding factor $(M_{200}/1.03\times 10^{15}\msun)^{1/6}$.

\begin{figure}
\includegraphics[width=\columnwidth]{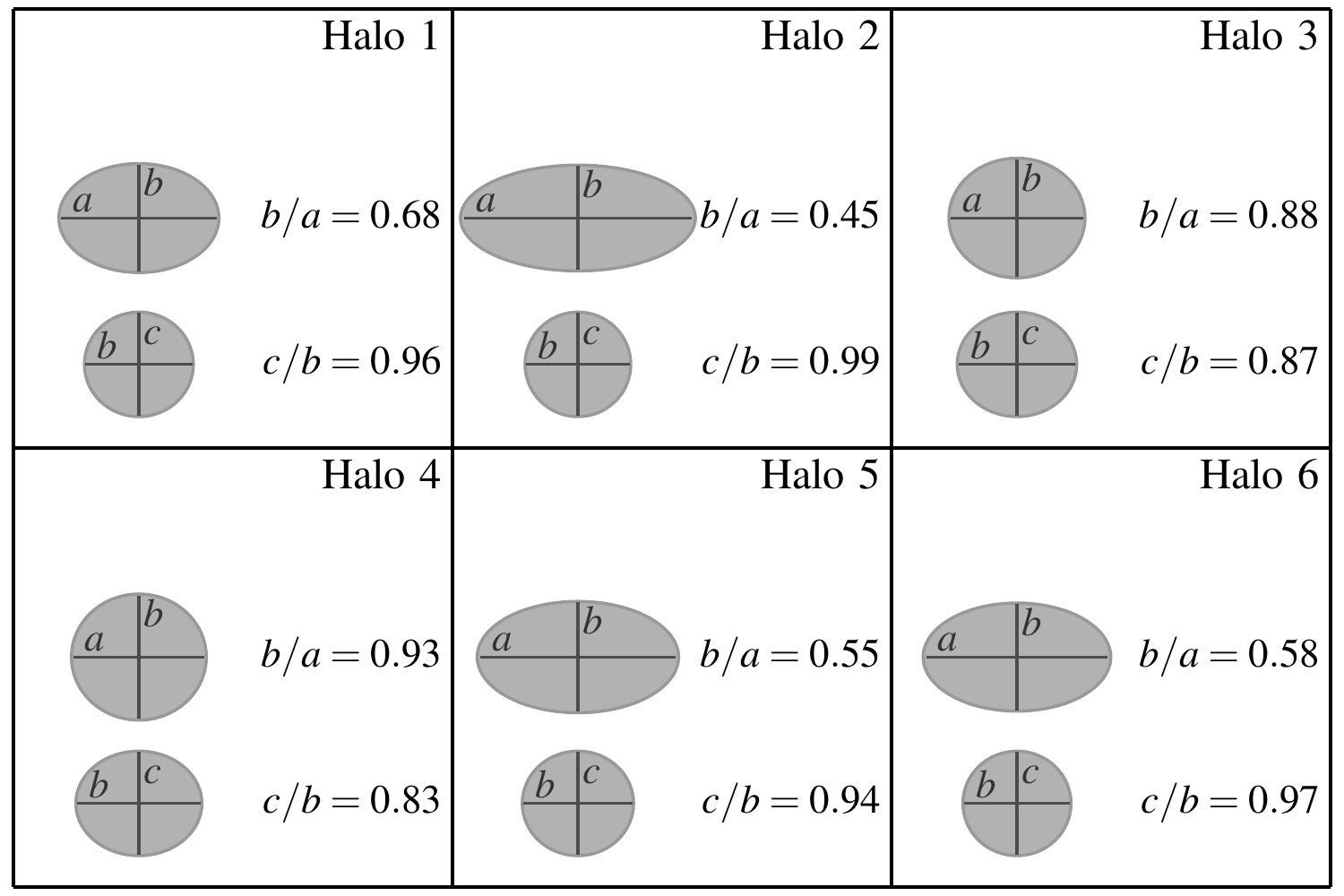}
\caption{The projection of the central stellar component of the six \eagle clusters
  along the minor axis (top ellipse in each panel) and major axis (bottom
  ellipse). The axes ratios are given next to each ellipse. Four of the galaxies are
  clearly prolate and the remaining two are slightly oblate. Prolate galaxies have
  velocity dispersions that are biased high (Fig.~\ref{fig:velocityDispersion}) when
  observed along their major axis.}
\label{fig:galaxyShape}
\end{figure}

The measured $\sigma_{\rm l.o.s.}$ is quite sensitive to the shape of the galaxy and
the viewing angle. The axial ratios, of the six \eagle BCGs (computed from the
principal axes of the inertia tensor of the star particles $a>b>c$) are illustrated
in Fig.~\ref{fig:galaxyShape} where the projection along the minor axis is shown at
the top of each panel and the projection along the major axis at the bottom.  Four of
the six \eagle clusters (1, 2, 5 and 6) are clearly prolate and the remaining two are
close to spherical.  We viewed the BCGs from 10,000 random directions placing an
imaginary slit at a random angle on the plane of the sky centred on the halo
potential minimum and measured the velocity dispersion of the stars as a function of
projected radius, subtracting any bulk rotation.

As expected, the line-of-sight velocity dispersions increase with radius. In the
inner regions ($r\lesssim10~\rm{kpc}$) gravity is dominated by the stars.  The
$1\sigma$ scatter from the different viewing angles, shown as a grey shaded region in
Fig.~\ref{fig:velocityDispersion}, is rather large at all radii for all objects, of
order 10\% or more for all but two of the halos.  The black solid line shows
$\sigma_{\rm{l.o.s.}}$ for a line-of-sight chosen along the major axis of each
BCG. In three of our six clusters (halos 2, 5 and 6) the velocity dispersion along
this particular line-of-sight is biased high and, in two cases, it falls outside the
$1\sigma$ scatter. As can be seen on Fig.~\ref{fig:galaxyShape}, these are the three
most prolate halos in our sample.  A bias in the line-of-sight velocity dispersion is
expected since orbits in prolate halos have larger velocities along the direction of
elongation.  These objects would nevertheless appear spherical on the sky when viewed
in this direction since the axis ratios $b/c$ are close to unity. The three most
spherical halos do not exhibit any particular bias when viewed along their major
axis, as expected.

The line-of-sight velocity dispersion profiles of the seven clusters studied by
\cite{Newman2013a} are shown as dashed colour lines with error bars in each panel of
Fig.~\ref{fig:velocityDispersion}. The six rescaled \eagle clusters have dispersions
that fall within the scatter of the observational data.  Thus, unless there is a
strong orientation bias for the BCGs in the cluster sample of \cite{Newman2013a}, a
mismatch in velocity dispersion profile cannot be the cause of the difference between
the slopes of the dark matter halos in the \eagle clusters and those inferred by
\citeauthor{Newman2013a}

Since the projected mass density of a prolate halo is also largest along its major
axis, there is a potential and well-understood selection bias in samples of clusters
selected for lensing studies. If the BCGs in the sample of \cite{Newman2013a} were
prolate and preferentially viewed along their major axes, then, as shown in
Fig.~~\ref{fig:velocityDispersion}, the observed line-of-sight velocity dispersions
would be biased high. This would lead to an overestimate of the mass enclosed within
the radius sampled by the velocity dispersion data. In the absence of other
information, it would not be possible to separate the relative contributions to this
estimate from stars and dark matter. However, the available lensing data constrains
the total mass (and, in the case of radial arcs, also the slope of the profile) in
the central regions of the cluster. This, together with the inferred stellar profile,
restricts the fits to the combined data and this could lead to an underestimate of
the dark matter mass near the centre of the clusters.

Such an effect could explain the difference between the slopes of the dark matter
profiles inferred by Newman et al. and those measured for the \eagle clusters.
However, Newman et al. argue that their sample does not suffer from such a bias since
the distribution of ellipticities in it is consistent with that of the BCG population
as a whole. In the case of A383, for which the X-ray data indicate is elongated along
the line-of-sight, they explicitly use a non-spherical model.

\subsection{Mass-to-light ratio}
\label{ssec:M_L}

As mentioned in Sec.~\ref{ssec:observed_slope}, the stellar mass density at a given
radius is degenerate with the dark matter mass density at that radius
(Eqn.~\ref{eq:measured_rho}). In the simulations we know the stellar mass and so we
can subtract it exactly from the total mass. The resulting value of the inner dark
matter halo slope was shown in Fig.~\ref{fig:slope_DM}. By contrast, in the
observational sample the stellar mass must be derived from an estimate of the stellar
mass-to-light ratio, $\Upsilon_*$.

\cite{Newman2013a,Newman2013b} treated $\Upsilon_*$ as a parameter in their model
over which they marginalize. The value of $\Upsilon_*$ is determined by the unknown
IMF; In their Bayesian model, they adopt a prior distribution that effectively
restricts the IMF to be between 1.5 times lighter than Chabrier and 2 times heavier
than Salpeter. Despite this wide range, the posterior distribution of $\Upsilon_*$ is
limited by the shape of the assumed prior in three out of the seven BCGs, suggesting
that the mass-to-light ratio is not well constrained by the data. Had we in our
simulations (which adopt a Chabrier IMF) subtracted a stellar mass inferred by
incorrectly assuming a Salpeter IMF, we would have overestimated $\Upsilon_*$ by
65\%. This would have led us to infer a negative slope for the inner dark matter
density profile in three out of our six clusters, implying virtually no dark matter
at their centres!

In Fig.~\ref{fig:slope_DM} we show the effect of overestimating $\Upsilon_*$ by a
much smaller factor of only 25\%.  The inferred slopes, shown by yellow stars, are
significantly shallower than the true slopes and have more scatter. Such a relatively
small systematic error would be sufficient to bring the inferred slopes in the
simulations into agreement with the estimates of \cite{Newman2013b}.

The estimate of $\Upsilon_*$ in the model of \cite{Newman2013b} requires the
measurement of the line-of-sight velocity dispersion profile, $\sigma_{\rm{l.o.s.}}$,
of the BCG, as a function of projected separation, $R$. In dynamical equilibrium,
$\sigma_{\rm{l.o.s.}}$ is given by the Jeans Equation \citep[e.g.][]{Binney1987,
  Cappellari2008}:
\begin{equation}
 \sigma_{\rm{l.o.s.}}^2(R) = \frac{2G}{\Sigma_*(R)}\int_R^\infty
 \frac{\mathcal{F}(r,R,\beta)\rho_*(r)M_{\rm{tot}}(r)}{r^{2-2\beta}}\rm{d}r,
 \label{eq:vel_disp}
\end{equation}
where $\rho_*(r)$ is the 3D density of tracers (the stars) whose surface density is
$\Sigma_*(R)$; $M_{\rm{tot}}(r)$ is the total enclosed mass;
$\beta=1-\sigma_\theta^2/\sigma_r^2$ is the velocity anisotropy parameter, here
assumed to be independent of radius, with $\sigma_r$ and $\sigma_\theta$ the radial
and tangential velocity dispersions, respectively\footnote{With this definition,
  $\beta=0, 1$ and $-\infty$ correspond to isotropic, radially biased and circularly
  biased orbits respectively.} and
\begin{align}
 \mathcal{F}(r,R,\beta) &=\frac{R^{1-2\beta}}{2} \left[\beta
   B\left(\frac{R^2}{r^2};\beta+\frac{1}{2},\frac{1}{2}\right) \right. \nonumber\\ &-
   \left.  B\left(\frac{R^2}{r^2};\beta-\frac{1}{2},\frac{1}{2}\right) +
   \frac{\sqrt{\pi}\left(3-2\beta\right)\Gamma(\beta-\frac{1}{2}
     )}{2\Gamma(\beta)}\right],\nonumber
\end{align}
where $\Gamma(x)$ is the Gamma function and $B(z;a,b)$ is the incomplete Beta
function. In the limit where $\beta\rightarrow0$, $\mathcal{F}(r,R,\beta)$ reduces to
\begin{equation}
 \lim_{\beta\rightarrow0}\mathcal{F}(r,R,\beta) = \sqrt{r^2 - R^2}. \nonumber
\end{equation}
In the more general case where $\beta$ is a function of $r$, the problem of
reconstructing the mass distribution becomes more complex.  Solutions for specific
forms of $\beta(r)$ have been derived by \cite{Mamon2010}.

\begin{figure}
\includegraphics[width=\columnwidth]{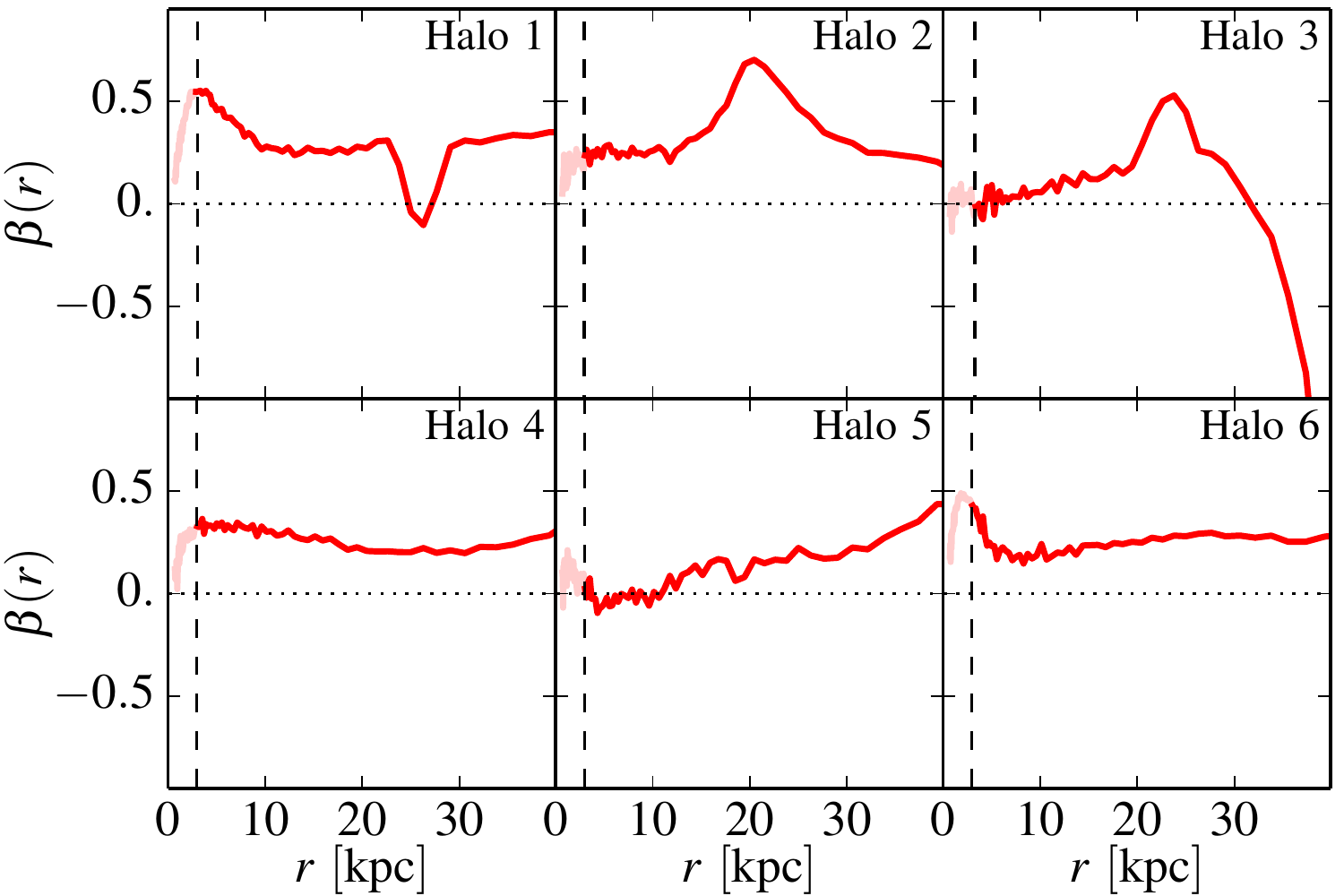}
\caption{Stellar anisotropy profile $\beta(r)$ as a function radius for the six
  \eagle clusters over the radial range relevant to the stellar kinematics analysis.
  The vertical dashed line on each panel shows the 3D convergence radius, $r_c$ and
  the profiles at lower radii are shown using shaded lines. Two BCGs are consistent
  with $\beta=0$ but would be better fit with a non-constant $\beta$.  Ignoring
  complex features, the other four clusters present more radially biased orbits with
  $\beta(r)\approx0.25$ A single profile shape for $\beta(r)$ cannot be used to
  characterize all six of our BCGs.}
\label{fig:anisotropy}
\end{figure}

In the Jeans equation the velocity anisotropy parameter and the mass are
degenerate. In their analysis \cite{Newman2013a} assumed $\beta=0$ i.e. isotropic
orbits. This assumption is a source of a potentially significant systematic error
which \citeauthor{Newman2013a} investigated. They found that if the orbits were
mildly radially biased with a constant value of $\beta=+0.2$, then $\Upsilon_*$ would
be overestimated by 30\%.  In our simulations we can calculate $\beta$ directly for
the stars in the model BCGs. The variation of $\beta$ with radius is shown in
Fig.~\ref{fig:anisotropy}. We find that, in general, $\beta$ varies with radius over
the range where Newman et al.  obtained kinematical data.  In two of our clusters,
$\beta$ is close to zero over this range, but in the other four, $\beta$ becomes
increasingly positive with radius, with a mean value of $\sim 0.2$ to $0.3$. Complex
features, which cannot be described by a simple linear form for $\beta(r)$ are also
present, precluding the reconstruction of $M(r)$ from an assumed functional form for
$\beta(r)$. It is also worth mentioning that the profile of $\beta(r)$ is
uncorrelated with the shape of the BCGs: of the two cases with nearly isotropic
orbits (halos 3 and 5), one is nearly spherical and the other very elongated (see
Fig.~\ref{fig:galaxyShape}).

In order to test the assumption of anisotropy, we inverted
eq.~\ref{eq:vel_disp} numerically. Extracting
$\sigma_{\rm{l.o.s.}}(R)$, $\Sigma_*(R)$ and $\rho_*(r)$ from the
simulated clusters we reconstructed $M_{\rm{tot}}(r)$ assuming
$\beta=0$ and compared the result to the actual value. We found that
for this assumption the reconstruction overestimates the value of
$M_{\rm{tot}}$ by factors ranging from 10\% to over 100\%.  Repeating
this analysis, this time assuming $\beta=0.2$, led to errors of
comparable size for the four halos that display an anisotropy profile
differing significantly from $\beta(r)=0.2$ (see Fig.
\ref{fig:anisotropy}).  Thus, for most of our clusters, the analysis
of \cite{Newman2013a} would have overestimated the stellar
mass-to-light ratio by more than the 25\% which, according to
Fig.~\ref{fig:slope_DM}, would reconcile their results with our
simulations.  This test, however, does not take into account
constraints on the \emph{total} mass profile from lensing data at
large radii, which could exaggerate the dependence of the inferred
value of $M_{\rm tot}$ near the centre on the assumed value of
$\beta$. 

In real clusters additional uncertainties are introduced by factors
such as an assumed form for the 3D stellar mass density profile,
$\rho_*(r)$, and an assumption for the value of the stellar
mass-to-light ratio, $\Upsilon_*$.  This is mitigated by constraints
on $M_{\rm tot}$ provided by lensing data although, in general, the
lensing and kinematical data do no overlap sufficiently to separate
the contributions from the stellar mass and the dark mass. In the
model of \citeauthor{Newman2013a}, $\Upsilon_*$ is coupled to other
parameters such as the slope of the total mass profile, so that the
effect on the quantity of interest, $\beta_{\rm DM}$, is difficult to
anticipate without re-running their pipeline for different assumptions
for the velocity anisotropy.  For example, \cite{Newman2013b} tried a
solution for the case of constant anisotropy, $\beta=0.2$, and found
an increase in $\beta_{\rm DM}$ of about $0.13$, which would bring
their data closer to our simulations.  What we can say with certainty
is that the kinematical model assumed by \citeauthor{Newman2013a} is
not consistent with the \eagle BCGs, offering a possible explanation
for the discrepancy in the dark matter density slopes.

Constraining the anisotropy, $\beta$, in cases in which, as in our simulated
clusters, it varies with radius is not straightforward. Yet, this is what is required
in order to lift the degeneracy between anisotropy and mass which lies behind the
degeneracy between $\Upsilon_*$ and the dark matter profile slope. The use of
Integral Field Spectroscopy may help constrain this quantity in future studies.

\section{Summary and conclusions}
\label{sec:summary}

We have studied the density profiles of the six most massive clusters in the largest
\eagle simulation \citep{Schaye2014}. The \eagle simulation was calibrated to provide
a good match to the observed stellar mass function and galaxy sizes in the local
universe, suggesting that it gives a realistic representation of the local galaxy
population.  Due to the relatively small volume of the simulation
($100^3~\rm{Mpc}^3$), the clusters selected for this study tend to be somewhat less
massive (mean $M_{200}= 2.6 \times 10^{14} \msun$) than the seven clusters studied by
\cite{Newman2013a} (mean $M_{200}=1 \times10^{15}\msun$) to which we compare our
results in particular detail, although the two lightest clusters in the observational
sample have similar masses to the three most massive \eagle clusters.  For these
clusters Newman et al. have obtained strong and weak lensing as well as stellar
kinematical data for the BCGs.

The total mass density profile of the \eagle clusters is dominated in the central
parts ($r<10~\rm{kpc}$) by the BCG. The presence of the central galaxy makes the
total mass profile steeper than an NFW profile near the centre.  The inner slope of
the total mass profile (defined as the average slope in the range $r=4-35~\rm{kpc}$)
agrees remarkably well with the slopes measured by \cite{Newman2013a} for their
clusters, with the corresponding slopes measured by \cite{Koopmans2009} for 58 early
type galaxies in the SLACS survey, and with the slope inferred by \cite{Agnello2014}
from the kinematics of globular clusters around M87.

The dark matter density profile of the \eagle clusters is very well described by the
NFW profile over the entire resolved radial range, $r=3-2000~\rm{kpc}$. By contrast,
\cite{Newman2013b}, after subtracting the contribution of the stars, inferred
significantly shallower dark matter slopes for their clusters in the inner regions,
in contradiction with our own results.  This discrepancy is puzzling because, in
addition to the total mass density profiles, the surface brightness and line-of-sight
velocity dispersion profiles of the \eagle clusters agree quite well with those of
the Newman et al. clusters.

We have considered possible explanations for the discrepancy between the inner dark
matter density profiles of the \eagle clusters and those inferred by
\cite{Newman2013b}. A possible interpretation is that the simulations lack the
correct physics, either because the dark matter is not collisionless
\citep[e.g.][]{Spergel2000, Vogelsberger2012,Rocha2013} or because extreme baryon
processes not represented in our simulations have destroyed the inner dark matter
cusps \citep{Martizzi2012}. Baryon effects associated with AGN in the \eagle
simulations are not strong enough to produce density cores; yet the simulation
reproduces the exponential cut-off in the stellar mass function remarkably well.

An alternative explanation for the discrepancy is that the uncertainties in the
determination of the inner dark matter density slope were underestimated by
\cite{Newman2013b}. In particular, their analysis relies on an accurate estimate of
the stellar mass-to-light ratios of the BCGs. We showed that a systematic
overestimation of this ratio by only 25\% would reconcile the observational data with
our results. An effect of this size could be produced if the measured stellar
velocity dispersions were biased high as would be the case if the BCGs (which are all
selected to be strong gravitational lenses) were prolate and preferentially viewed
along their major axis. However, \cite{Newman2013a,Newman2013b} have argued that such
a selection bias is unlikely in their sample since the distribution of BCG
ellipticities appears to be typical of a randomly oriented population.

Another possible source of systematic error in the estimate of the
stellar mass-to-light ratio is the assumption made by
\cite{Newman2013b} that the stars in the BCG have a uniform and
isotropic distribution of orbits. In their paper, they showed that
mildly radial orbits would lead to an overestimate of the stellar
mass-to-light ratio of 30\%, sufficient, in principle, to account for
the discrepancy with the NFW inner dark matter slopes of the \eagle
clusters. We find that just such a situation is present in four of our
six clusters which show radially biased orbital distributions which
vary with radius in a complicated way. However, in practice, the
situation is not straightforward because the mass-to-light ratio in
the model of \citeauthor{Newman2013a} is coupled to other parameters
and is sensitive to the constraints on the total mass profile from
lensing.

We can conclude, however, that systematic errors resulting from the
assumptions made in the analysis of \cite{Newman2013b} could
potentially be large enough to account for the shallow inner dark
matter profiles that these authors infer for their clusters, in
conflict with the cuspy profiles found for the \eagle clusters.
Unfortunately it is very difficult, if not impossible, to break the
degeneracies inherent in stellar kinematical analyses with existing
data. High resolution integral field spectroscopy of BCGs could prove
helpful in future work.

\section*{Acknowledgements}
We thank Lydia Heck and Peter Draper for their indispensable technical assistance and
support.  We also thank for Drew Newman and Tomasso Treu for comments on an early
draft of this paper which influenced its final form.  RAC is a Royal Society
University Research Fellow. This work was supported by the Science and Technology
Facilities Council (grant number ST/F001166/1); by the Dutch National Computing
Facilities Foundation (NCF) for the use of supercomputer facilities, with financial
support from the Netherlands Organization for Scientific Research (NWO); the European
Research Council under the European Union's Seventh Framework Programme
(FP7/2007-2013) / ERC Grant agreements 278594-GasAroundGalaxies, GA 267291 Cosmiway,
GA 238356 Cosmocomp and the Interuniversity Attraction Poles Programme initiated by
the Belgian Science Policy Office ([AP P7/08 CHARM]).  It used the DiRAC Data Centric
system at Durham University, operated by the Institute for Computational Cosmology on
behalf of the STFC DiRAC HPC Facility (www.dirac.ac.uk). This equipment was funded by
BIS National E-infrastructure capital grant ST/K00042X/1, STFC capital grant
ST/H008519/1, and STFC DiRAC Operations grant ST/K003267/1 and Durham
University. DiRAC is part of the National E-Infrastructure. We acknowledge PRACE for
awarding resources in the Curie machine at TGCC, CEA, Bruy\`eres-le-Ch\^atel, France.

\bibliographystyle{mnras} \bibliography{./bibliography.bib}

\label{lastpage}

\end{document}